\documentclass[aps,prl,twocolumn,groupedaddress]{revtex4-1}
\usepackage[mathscr]{eucal}
\usepackage{color}
\usepackage{graphicx}
\usepackage{graphics}
\usepackage{amsmath}
\usepackage{amssymb}
\usepackage{anysize}
\usepackage{caption}
\usepackage{subcaption}
\usepackage{multirow}
\usepackage{pbox}

\begin{document}


\title{Connecting electron and phonon spectroscopy data to consistently determine quasiparticle-phonon coupling on the surface of topological insulators}


\author{Colin Howard and M. El-Batanouny}
\affiliation{Department of Physics, Boston University, Boston, Massachusetts 02215, USA}


\date{\today}

\begin{abstract}
Photoemission and phonon spectroscopies have yielded widely varying estimates of the electron-phonon coupling constant $\lambda$ on the surfaces of topological insulators, even for a particular material and technique. We connect the results of these experiments by determining the Dirac fermion quasiparticle spectral function using information from measured spectra of a strongly-interacting, low-lying optical surface phonon band. The manifest spectral features resulting from the coupling are found to vary on energy scales $< 1$ meV, and are distinct from those traditionally observed in the case of acoustic phonons in metals. We explore different means of determining $\lambda$ from the electron perspective and identify definitions that yield values consistent with phonon spectroscopy.
\end{abstract}


\maketitle


The new class of materials coined topological insulators (TIs)\cite{Fu3,Hasan,Hasan2,Qi} has been the subject of extensive studies over the past five years and continues to be one of the most active research areas of condensed matter physics. The main attraction in studying these materials, from both fundamental and technological perspectives, has been the presence of chiral Dirac fermion quasiparticles (DFQs) that define a robust metallic surface state protected against backscattering by time-reversal symmetry. Most of the reported studies adopted a non-interacting, single-particle approach to the DFQ system\cite{Zhang2,Zhang1,Yazyev}. However, recently, there have been several experimental and theoretical reports on the interaction of the DFQs with surface phonons. Currently, there is little consensus about the magnitude of this coupling as evidenced by widely varying values of the electron phonon coupling (epc) parameter $\lambda$ appearing in the literature. 

We begin by noting that we have reported experimental surface phonon dispersion measurements for the 3D TIs Bi$_2$Se$_3$\cite{zhu} and Bi$_2$Te$_3$\cite{Howard} using helium atom-surface scattering (HASS). Our results revealed the absence of acoustic Rayleigh modes and the presence of strong Kohn anomalies arising from strong coupling of DFQs to low-lying surface optical phonon modes \cite{Zhu2, Howard}, yielding $\lambda = 0.7$ and $\lambda = 2.0$ for Bi$_2$Se$_3$ and Bi$_2$Te$_3$, respectively\footnote{The values quoted here are averages over the entire 2D surface Brillouin zone, making them larger than the simple average of $\lambda({\bf q})$ along a particular high-symmetry direction}. Three theoretical studies\cite{Thalmeier,Giraud,Sarma} considered the interaction of DFQs with long-wavelength surface acoustic modes. In all three the strength of the epc was found to be quite small, which is actually consistent with our results. As a matter of fact, they justify the absence of acoustic Rayleigh phonons in HASS data: It is well established that the thermal energy helium atoms employed in HASS are scattered by the surface electron density about 2-3 \AA\ above the terminal layer of atomic nuclei. Thus, detection of surface phonons by HASS involves scattering from the phonon-induced surface electron density oscillations. The results of the three theoretical studies confirm that surface acoustic phonons are weakly coupled to the surface metallic charge-density (DFQs) so that the induced density oscillations are effectively suppressed. 

In light of this it appears that the strong epc is indeed the result of interactions with low-energy optical phonons on the surface, with the acoustic modes playing little role. Indeed, we call attention to the fact that the metallic DFQ surface state in TIs is actually more akin to a 2D electron gas in semiconductor heterojunctions and quantum wells than to a 2D metal. In such systems the dominant interaction of 2D metallic electrons is with long-wavelength optical phonons of the host semiconductor\cite{Sarma2} via Fr\"olich type coupling, not acoustic phonons. By analogy, the dominant interaction of the DFQs is with the optical phonons of the host TI. Moreover, unlike metallic surfaces, where the typical ratio of $E_F$ to phonon energies is $\sim10^3$, the Fermi energy of TIs is $\sim 100$ meV while phonon energies are $\sim 10$ meV.

Other experimental estimates of the DFQ-phonon coupling parameter $\lambda$ have been provided by angle-resolved photoemission spectroscopy (ARPES). Early estimates for Bi$_2$Se$_3$ were obtained by performing a linear fit to the imaginary part of the DFQ self energy. The reported values were: $\lambda=0.25$\cite{Hofmann2} ($\Delta E< 15$ meV energy resolution) and $\lambda=0.08$\cite{Valla} ($\Delta E\sim 8$-$12$ meV). However, more recent studies employing high-resolution ARPES instruments, $\Delta E\sim 1$ meV, revealed more detailed texture in the DFQ dispersion curve near the Fermi surface \cite{Kondo,Chen3}. Kondo and coworkers reported a surprisingly large $\lambda\sim 3$ for Bi$_2$Se$_3$ and Bi$_2$Te$_3$, which they attributed to contributions from phonon as well as spin/plasmon\cite{Zhang4} type interactions. In contrast Chen {\it et al} reported a significantly lower $\lambda = 0.19$ for  Bi$_2$Te$_3$. 
   
In this letter we attempt to clarify the sources of discrepancy among the experimental results and define a consistent approach to investigating the DFQ-phonon interactions. To this end, we connect results of phonon dispersion and ARPES measurements by presenting DFQ spectral functions (SFs) derived from a Matsubara Green function containing the optical phonon dispersion for Bi$_2$Te$_3$ together with the associated epc matrix elements\cite{Howard}. We establish similarities between the DFQ dispersion and linewidths obtained from our calculated SF and the recent high-resolution ARPES data in the relevant energy range. We demonstrate, based on these similarities, that the major contribution to $\lambda$ comes from the low-lying surface optical phonon band. Furthermore, our calculated SFs show that high instrument resolution and cryogenic sample temperatures are necessary to probe the energy scales associated with the epc. We also show that it is possible to obtain consistent values of $\lambda$ from both electron and phonon perspectives. 

We start with the DFQ finite-temperature Matsubara Green function 

\small 
\begin{align}
&G({\bf k},\tau,T)=-\frac{1}{{\cal A}}\sum_{{\bf k}_1{\bf k}_2\atop{\bf q}}\;\left|\text{g}_{\bf q}\right|^2\iint_0^\beta\;d\tau_1\,d\tau_2\;\mathscr{D}({\bf q},\tau_1-\tau_2) \notag \\
&\times\left<T_\tau\left[c^\dagger_{{\bf k}_1+{\bf q}}(\tau_1)c^\dagger_{{\bf k}_2-{\bf q}}(\tau_2)c_{{\bf k}_2}(\tau_2)c_{{\bf k}_1}(\tau_1)c_{\bf k}(\tau)c^\dagger_{\bf k}(0)\right]\right>
\end{align}
\normalsize 
where ${\cal A}$ is the surface area, $\text{g}_{\bf q}$ is the electron-phonon matrix element, $\mathscr{D}({\bf q},\tau_1-\tau_2)$ is the phonon propagator, $c_{\bf k}^\dagger,\;c_{\bf k}$ are the electron creation and annihilation operators, respectively and $T_\tau$ is the imaginary-time ordering operator. We neglect the weak direct Coulomb interactions in the DFQ system. This is warranted by the fact that TIs possess large dielectric constants ($\kappa>50$) and Fermi velocities  $\sim10^5$ m/s that yield a small effective fine-structure constant $\alpha = e^2/(\kappa\hbar v_F )\approx 0.05$\cite{Zhang4,Sarma}. Moreover, from a Fermi liquid perspective, the quasiparticle nature of the DFQs close to the Fermi energy ($E_F$) is well defined. Since our analysis will focus on a region $\pm 7$ meV about $E_F$, the direct electron-electron interactions need not be considered. 

Fourier transforming the Matsubara function gives the DFQ self-energy 

\small
\begin{align}\label{sigma}
&\Sigma({\bf k},i\omega_n,T)\,=\,\frac{1}{{\cal A}}\ \sum_{\bf q}\ \left|\text{g}_{\bf q}\right|^2\;\Bigl(1+\hat{\bf k}\boldsymbol{\cdot}\widehat{{\bf k}+{\bf q}}\Bigr) \notag \\
&\times\left[\frac{n_B(\omega_{\bf q})+n_F(\varepsilon_{{\bf k}+{\bf q}})}{i\omega_n-\varepsilon_{{\bf k}+{\bf q}}+\omega_{\bf q}}+\frac{n_B(\omega_{\bf q})+1-n_F(\varepsilon_{{\bf k}+{\bf q}})}{i\omega_n-\varepsilon_{{\bf k}+{\bf q}}-\omega_{\bf q}}\right]
\end{align} 
\normalsize
where $\omega_n$ is the Matsubara frequency, and $\omega_{\bf q}$ is the optical phonon frequency at wavevector ${\bf q}$. We shall use the dispersion curve and phonon matrix element $\text{g}_{\bf q}$ obtained by fitting the data in \cite{Howard} for Bi$_2$Te$_3$. Analytically continuing and replacing the sum by an integral, we obtain

\small
\begin{align}
&\Sigma({\bf k},\omega,T)=\frac{1}{(2\pi)^2}\int d{\bf q}\ \left|\text{g}_{\bf q}\right|^2\Bigl(1+\hat{\bf k}\boldsymbol{\cdot}\widehat{{\bf k}+{\bf q}}\Bigr)\notag \\ 
&\times\left[\frac{n_B(\omega_{\bf q})+n_F(\varepsilon_{{\bf k}+{\bf q}})}{\omega-\varepsilon_{{\bf k}+{\bf q}}+\omega_{\bf q}+i\delta} +\frac{n_B(\omega_{\bf q})+1-n_F(\varepsilon_{{\bf k}+{\bf q}})}{\omega-\varepsilon_{{\bf k}+{\bf q}}-\omega_{\bf q}+i\delta}\right]
\end{align} 
\normalsize
with the imaginary part given by

\small
\begin{align}
&Im[\Sigma({\bf k},\omega,T)]=-\frac{1}{4\pi}\int dq \,q\, \int d\varphi \left|\text{g}_{\bf q}\right|^2 \Bigl(1+\hat{\bf k}\boldsymbol{\cdot}\widehat{{\bf k}+{\bf q}}\Bigr)\notag\\ 
&\times\left[\Bigl(n_B(\omega_{\bf q})+n_F(\varepsilon_{{\bf k}+{\bf q}})\Bigr)\delta(\omega-\varepsilon_{{\bf k}+{\bf q}}+\omega_{\bf q})\right. \notag \\
&\left.+\Bigl(n_B(\omega_{\bf q})+1-n_F(\varepsilon_{{\bf k}+{\bf q}})\Bigr)\delta(\omega-\varepsilon_{{\bf k}+{\bf q}}-\omega_{\bf q})\right]
\end{align} 
\normalsize
We first determine $Im\,[\Sigma({\bf k},\omega,T)]$ numerically and then obtain $Re\,[\Sigma({\bf k},\omega,T)]$ via the Kramers-Kronig relations. Finally, we obtain the DFQ SF as
\begin{widetext}
\begin{equation}
A({\bf k},\omega,T)\,=\,\frac{1}{\pi}\frac{\left| Im\,[\Sigma({\bf k},\omega,T)]\right|}{\left(\omega-\hbar\,v_F(|{\bf k}|-k_F)-Re\,[\Sigma({\bf k},\omega,T)]\right)^2+Im\,[\Sigma({\bf k},\omega,T)]^2} 
\end{equation}
\end{widetext}
where $\omega$ is measured from $E_F = \hbar v_Fk_F$ and we've used $\varepsilon_{\bf k} = \hbar\,v_F(|{\bf k}|-k_F)$ for the nominal dispersion of the DFQs above the Dirac point. The determination of $A({\bf k},\omega,T)$ allows direct comparison of the quasiparticle energy dispersion and state broadening with experimental results obtained by ARPES measurements. Moreover, the density of DFQ states
\begin{align}
{\cal N}(\omega,T)\,=\,\frac{1}{k_F}\int\;dk\,A({\bf k},\omega,T)
\end{align}
allows for comparison with experimental results of ARPES energy distribution curves (EDCs) and scanning tunneling spectroscopy (STS).

\begin{figure*}
\hspace{-0.65in}
\begin{subfigure}[b]{0.289\linewidth}
	\includegraphics[scale=0.37]{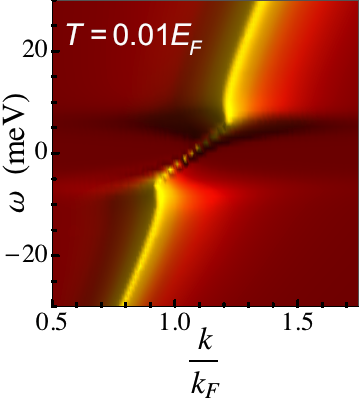}
\end{subfigure}
\begin{subfigure}[b]{.247\linewidth}
	\includegraphics[scale=0.37]{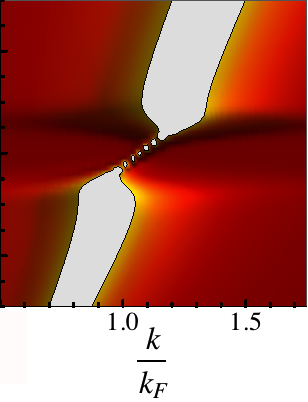}
\end{subfigure}
\begin{subfigure}[b]{.248\linewidth}
	\includegraphics[scale=0.37]{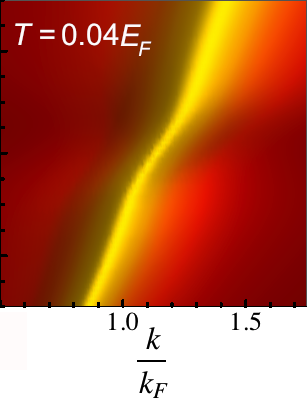}
\end{subfigure}
\begin{subfigure}[b]{.18\linewidth}
	\includegraphics[scale=0.37]{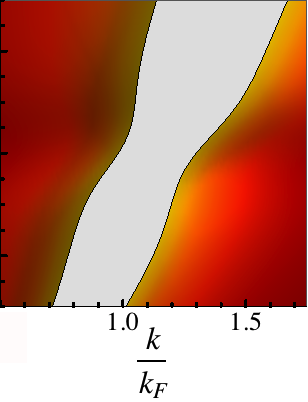}
\end{subfigure}
\caption{Calculated spectral function and momentum distribution curve FWHM $\Delta k(\omega)$ for Bi$_2$Te$_3$ at $T = 0.01E_F$ and $T = 0.04E_F$. The white space in the second and fourth figures is used to indicate $\Delta k(\omega)$. \label{fig:SF}}
\end{figure*}

Since reported high-resolution ARPES results were performed at temperatures $T\sim7$-$20$ K, we shall focus our analysis at similarly low temperatures. The calculated SF for $T=0.01E_F\equiv T_1$ is presented in the left panel of figure \ref{fig:SF} where the DFQ band dispersion appears as the bright curve. The epc footprint is readily apparent as their deviation from the linear dispersive behavior within $\pm7$ meV of $E_F$, where two kinks appear, one slightly above and the other below $E_F$, pointing to large velocity renormalization. We note that the structural details of the kinks are discernible on an energy scale $< 1$ meV. We find that both of these energy scales are quite small compared to the ARPES resolutions quoted in Refs. \cite{Hofmann2, Valla}, which could account for the fact that no such deviation in the dispersion nor significant temperature dependence of $Im[\Sigma]$ was observed in those experiments. The application of higher resolution in \cite{Kondo,Chen3} brought some of these features to light.

The second panel from the left in figure \ref{fig:SF} shows $\Delta k(\omega)$, the full-width-half-maximum (FWHM) of the momentum distribution curves (MDCs), as the white region. Note that $\Delta k(\omega)$ increases from $0.2k_F$ at $\omega=-30$ meV to $0.28k_F$ at $\omega=-7$ meV, where the lower kink occurs, indicating a gradual increase in the strength of the epc. The important observation is that $\Delta k(\omega)$ abruptly shrinks, reaching negligible values above $\omega=-2$ meV, and resumes a linear dispersion but with a slower velocity. The collapse in the peak width signals the absence of DFQ coupling to phonons. This termination of the coupling is consistent with the observation that the DFQs interact strongly with low-lying optical phonon modes, whose lower band-edge occurs at $~3$ meV, as reported in HASS data. Unlike acoustic phonons, the interaction does not extend to infinitesimal energies close to $E_F$, which is why $\Delta k(\omega)$ is suppressed in the region $\sim \pm 2$ meV. 

We determine $\lambda$ at $T_1$ in two distinct, albeit equivalent ways. In the first method, we apply the definition\cite{Grimvall,Hofmann}
\begin{equation}
\lambda=-\frac{\partial Re\,[\Sigma]}{\partial\omega}\Bigl|_{\omega=E_F}
\label{eq:lam1}
\end{equation}
that has traditionally been used for metallic surfaces. The alternative approach utilizes the relation 
\begin{equation}
\lambda = \frac{v_0}{v_F}-1
\label{eq:lam2}
\end{equation}  
where $v_F$ is the epc renormalized Fermi velocity and $v_0$ is the un-renormalized value. Both definitions yield $\lambda = 2$, which is consistent with results from phonon spectroscopy.  

Equation \ref{eq:lam2} evokes an analogy between the relativistic DFQs propagating on a TI surface and light traveling in a dielectric medium, which helps illuminate the physical meaning of $\lambda$. First, we note that the nominal linear dispersion of the DFQs reflects their massless character, invalidating the application of the conventional idea of mass enhancement as defined by $m*/m = 1+\lambda$ to TIs. Instead, it is appropriate to interpret $\lambda$ as a velocity renormalization factor via $v_0/v_F = 1 + \lambda$. We note that this is totally consistent with, and actually more fundamental than the previous relation, to which it simplifies when applied to parabolic energy bands. Thus, $\lambda$ provides a measure of the renormalization of the group velocity of the DFQs near the Fermi energy, much like the index of refraction does for light in a dielectric medium. Just as light slows down when propagating through matter, the DFQs near the Fermi energy are slowed by their interactions with the phonon gas.   

At this point it is appropriate to compare our calculated SF with the recent high-resolution ARPES data. The authors of Ref. \cite{Kondo} identify coupling between DFQs and two distinct bosonic excitations as evidenced by deviations from linear dispersion and modifications to $\Delta k(\omega)$ in the vicinity of $-3$ meV and $\sim-15\to -20$ meV. They attribute the latter to coupling to a high-energy phonon mode which we do not consider in our analysis, partly because such strong coupling had no clear footprint in our HASS data. The exclusion of this coupling in our calculations leads to some disagreements in the region $\omega<-10$ meV. For example our $\Delta k(\omega)$ decreases with decreasing $\omega$ in the range $-10\to-30$ meV, whereas the high resolution ARPES data exhibits an increase. However, our main focus here is on the energy range $0\le\omega>-7$ meV, especially the feature at $-3$ meV which the authors attribute to coupling to both spin/plasmon excitations and the low-lying optical surface phonon band identified in our HASS data. This assignment was deemed necessary by the authors to account for the large value of $\lambda\sim3$ they obtained. Yet, our analysis above shows that a dominant contribution of $\lambda=2$ comes from the surface optical phonon band. We speculate that coupling to the higher energy phonon band may be sufficient to account for the difference, without the need to invoke additional contributions from spin/plasmon interactions. The second high-resolution ARPES study\cite{Chen3} estimate of $\lambda=0.19$ is much smaller than that determined from our calculations and the estimate in Ref. \cite{Kondo}. It appears the authors overlooked the presence of the $3$ meV peak in their spectra, resulting in an underestimate of $\lambda$ when applying equation \ref{eq:lam1}.   
 
\renewcommand{\arraystretch}{2.0}

\begin{table}
\begin{center}
\caption{Experimental estimates of the epc constant. Values labeled (A),(B),(C),(D) appear in Refs. \cite{Hofmann2},\cite{Valla},\cite{Kondo},\cite{Chen3} respectively.}
\label{lambdatabel}
\begin{tabular}{| l | l | l | l |}
\hline
System& $\lambda$ & \pbox{20 cm}{Resolution \\ \hspace{-0.25 in}(meV)} & Method\\
\hline\hline
\multirow{2}{*}{Bi$_2$Se$_3$} & 0.25 (A) & $< 15$ & \multirow{2}{*}{$\displaystyle{\frac{1}{\pi k_B}\frac{\partial Im[\Sigma]}{\partial T}}$}\\
\cline{2-3}
 & 0.08 (B) & $8-12$  & \\
\hline
Bi$_2$Se$_3$, Bi$_2$Te$_3$ & 3 (C) & 1  & $\displaystyle{\frac{v_0}{v_F} - 1}$\\[5 pt]
\hline
Bi$_2$Te$_3$ & 0.19 (D) & 1  & $\displaystyle{-\frac{\partial Re[\Sigma]}{\partial\omega}\Bigl|_{\omega=E_F}}$\\[5 pt]
\hline
\hline
\multirow{3}{*}{Bi$_2$Te$_3$} & \multirow{2}{*}{2*} & \multirow{2}{*}{1}  & $\displaystyle{-\frac{\partial Re[\Sigma]}{\partial\omega}\Bigl|_{\omega=E_F}}$ \\[5 pt]
\cline{4-4}
& & &$\displaystyle{\frac{v_0}{v_F}-1}$\\[5 pt]
\cline{2-4}
& 2   & 1  &  HASS \\
\hline
\end{tabular}
\end{center}
\raggedright{\it{*Value obtained from current calculations}}
\end{table}

For the sake of completeness, we studied the effects of increased temperature by calculating the SF at $T=0.04E_F\equiv T_2$. The results are plotted in the last two panels of figure \ref{fig:SF}. One can discern dramatic differences between the SFs at the two temperatures. First, the $T_2$ SF displays enhanced line broadening as compared with the $T_1$ SF. More importantly, it also exhibits a much smaller deviation from the nominal linear dispersion compared to its $T_1$ counterpart, emphasizing the fact that high resolution and cryogenic temperatures are required for adequate observation of the epc manifest features in ARPES measurements. Indeed, we find a significantly higher $v_F$ at $T_2$, yielding a markedly lower value of $\lambda$ when applying equation \ref{eq:lam2}. This casts doubt on the applicability of a linear temperature dependence of $Im[\Sigma]$ to extract an estimate of $\lambda$, since this method assumes it is constant over the temperature range of interest.  

\begin{figure}
\includegraphics[width=1.0\linewidth]{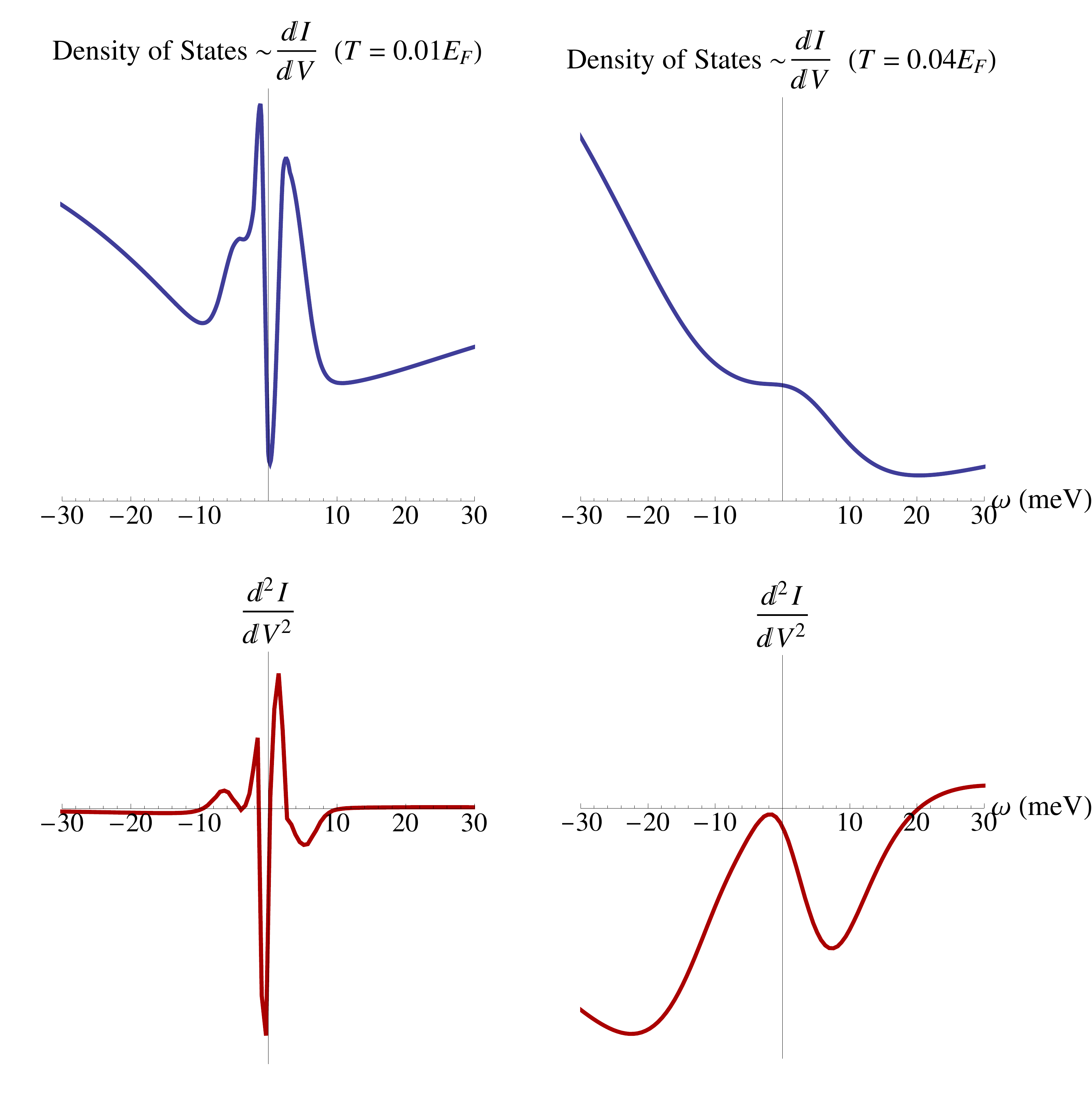}
\caption{\label{fig:DoS}The density of states and its derivative obtained from the calculated spectral function at $T = 0.01E_F$ (left) and $T=0.04E_F$ (right).} 
\end{figure}

Another noteworthy feature in our results is the DFQ density of states (DoS) and its derivative, which we present for $T_1$ in the left side of figure \ref{fig:DoS}. Manifestations of the epc below $E_F$ appear as a peak-dip-hump structure at $(-2)$-$(-3)$-$(-4)$ meV, respectively, in agreement with \cite{Kondo}. Additionally, peaks in $d^2I/dV^2\equiv \partial{\cal N}/\partial\omega$ appear at $-2$ meV and $-7$ meV which are consistent with STS measurements by Madhavan's group \cite{Madhavan}. We note that these features are completely absent in the $T_2$ data, where the increased temperature has smoothed away the kinks in the SF and abrupt changes in $\Delta k(\omega)$. The readily apparent texture in the $T_1$ calculation suggests that STS is a valuable tool for observing signatures of epc on the surfaces of TIs.

In this letter we have established the consistency between theoretical predictions of negligible DFQ coupling to surface acoustic phonons and the apparent absence of Rayleigh phonon modes in HASS measurements. Using the dispersion and epc matrix elements of the optical phonon band that couples strongly to DFQs, we calculated the corresponding SF in the vicinity of the Fermi energy on the surface of Bi$_2$Te$_3$. We identified the manifest features of the epc in the SF, thereby completing the translation from the phonon to the electron perspective. Our results indicate that the SF undergoes significant renormalization beginning at $-7$ meV where the dispersion develops a kink before continuing linearly upward, for $\omega>-2$ meV, with diminished group velocity. Additionally, the MDC  linewidth $\Delta k(\omega)$ undergoes growth in the region $-7 \to -3$ meV indicating increased coupling, but collapses suddenly at  approximately $-2$ meV due to the lower energy cutoff of the interacting optical phonon band. We calculate a value $\lambda=2$ consistent with HASS measurements but acknowledge that additional coupling to higher energy phonon bands could increase this value. It is our hope that the considerations presented here have unveiled the consistency in what initially appeared to be disparate theoretical and experimental results. 

We would like to thank Claudio Chamon for very useful discussions regarding this manuscript. We also extend our thanks to Vidya Madhavan for providing valuable insight into the nature of STS spectra on the surfaces of TIs.    

\bibliography{refs}
\end{document}